\documentclass[sn-mathphys-num]{sn-jnl}

\usepackage[utf8]{inputenc}
\usepackage[T1]{fontenc}
\usepackage{amsmath}
\usepackage{amsfonts}
\usepackage{amssymb}
\usepackage{stmaryrd}
\usepackage{graphicx}

\usepackage{listings}
\usepackage{xcolor}

\begin{document}

\title[Assessing GPT Model Uncertainty in Mathematical OCR Tasks via  Entropy Analysis]{Assessing GPT Model Uncertainty in Mathematical OCR Tasks via  Entropy Analysis}

\author*{\fnm{Alexei} \sur{Kaltchenko}}\email{akaltchenko@wlu.ca}

\affil*{\orgdiv{Physics and Computer Science}, \orgname{Wilfrid Laurier University}, \orgaddress{\street{75~University~Ave~W}, \city{Waterloo}, \postcode{N2L3C5}, \state{Ontario}, \country{Canada}}}

\abstract{This paper investigates the uncertainty of Generative Pre-trained Transformer (GPT) models in extracting mathematical equations from images of varying resolutions and converting them into LaTeX code. We employ concepts of entropy and mutual information to examine the recognition process and assess the model's uncertainty in this Optical Character Recognition (OCR) task. By analyzing the conditional entropy of the output token sequences, we provide both theoretical insights and practical measurements of the GPT model's performance given different image qualities.

Our experimental results, obtained using a Python implementation available on GitHub, demonstrate a clear relationship between image resolution and GPT model uncertainty. Higher-resolution images lead to lower entropy values, indicating reduced uncertainty and improved accuracy in the recognized LaTeX code. Conversely, lower-resolution images result in increased entropy, reflecting higher uncertainty and a higher likelihood of recognition errors. These findings highlight the practical importance of considering image quality in GPT-based mathematical OCR applications and demonstrate how entropy analysis, grounded in information-theoretic concepts, can effectively quantify model uncertainty in real-world tasks.}


\maketitle

\section{Introduction and Literature Review}

Uncertainty quantification (UQ) in large language models (LLMs), such as Generative Pre-trained Transformer (GPT) models, has become increasingly important as these models are deployed in applications where reliability and trustworthiness are critical. One such application is Optical Character Recognition (OCR), particularly in tasks involving the recognition and extraction of mathematical expressions from images. The complexity of mathematical notation and the variability introduced by different image resolutions present unique challenges in this domain.

Early work in deep learning UQ provided a foundation for understanding and reducing uncertainties in machine learning models by exploring methods like Bayesian approximation and ensemble learning~\cite{f14fc9e399d44463a17cc47a9b339b58f6ef7502}. These methods aimed to enhance model performance across various fields, including computer vision and natural language processing.

In the context of language models, initial studies demonstrated that larger models begin to learn tasks without explicit supervision, suggesting inherent abilities in models like GPT-2 to generalize across tasks~\cite{9405cc0d6169988371b2755e573cc28650d14dfe}. However, the need for models to assess the validity of their own outputs led researchers to investigate self-evaluation mechanisms. For instance, it was shown that LLMs could predict the correctness of their answers and evaluate the probability of their responses being true, improving calibration and performance across diverse tasks~\cite{142ebbf4760145f591166bde2564ac70c001e927}.

Advancements continued with efforts to enable GPT-3 models to express uncertainty in natural language, bridging the gap between probabilistic confidence and verbalized expressions of certainty~\cite{374dd173491a59a10bbb2b3519ebcfe3649f529d}. This approach allowed models to generate answers alongside confidence levels (e.g., "90\% confidence"), providing more transparent and calibrated responses.

Given the increasing prevalence of closed-source LLMs and limitations in accessing internal model information, black-box methods for UQ have gained attention. Researchers have developed techniques to quantify uncertainty without requiring modifications to models, such as measuring semantic dispersion to predict response quality~\cite{ad934a9344f68fcc0b9aa704102aa48c39c5b591}. Studies have evaluated confidence elicitation strategies in black-box LLMs, highlighting challenges in calibration and overconfidence, and suggesting methods to mitigate these issues through prompting strategies and consistency measures~\cite{8f7297454d7f44365b9bcda5ebb9439a43daf5e6}. Alternative approaches include estimating uncertainty via multiple rephrased queries~\cite{6dfb33f4229852d187d12937e6daaf01fb7bed77} and training interpretable models on engineered features to estimate confidence~\cite{306abfd0fec0cfe1b2c3960648a25f2d9efbe26a}.

A significant line of research has explored the use of entropy as a measure of uncertainty in LLMs. Entropy, a fundamental concept in information theory, quantifies the unpredictability of a system. In early applications outside natural language processing, entropy was used to measure uncertainties in geological models, providing quantitative insights into model indefiniteness and data assimilation~\cite{e53ab119b699d24a1f1a51689c012676fcd8867b}. Similarly, entropy-based uncertainty measures were applied in feature selection for gene expression data classification, handling noise and improving predictive accuracy~\cite{6ca2a7813df96c4c93908bda183f819d35d5d6ee}.

In the realm of LLMs, semantic entropy has been introduced to estimate uncertainty in natural language generation tasks. By accounting for linguistic invariances and shared meanings, semantic entropy provides a more predictive measure of model accuracy, overcoming challenges posed by semantic equivalence in different sentences~\cite{507465f8d46489a68a527cb5304d76bdb6c31ed9}. Building upon this concept, Semantic Entropy Probes (SEPs) have been proposed as robust and efficient methods for hallucination detection in LLMs. SEPs approximate semantic entropy directly from hidden states of a single generation, reducing computational overhead while retaining high performance in uncertainty quantification~\cite{648375ec8d90cb792de76030223539498612102e}.

Further extending entropy-based methods, the notion of semantic density has been introduced to extract uncertainty information from probability distributions in semantic space. This approach addresses limitations of previous methods by being task-agnostic and not requiring additional training or data. Experiments have demonstrated superior performance and robustness of semantic density in uncertainty quantification across various state-of-the-art LLMs and benchmarks~\cite{57269ec0822fb0afcb16a53558dbcd8ebe9a0a7d}.

Other studies have connected GPT models with information-theoretic concepts like Kolmogorov complexity. By utilizing GPT-based compression, researchers have approximated information distance for few-shot learning, justifying pre-training objectives of GPT models and enhancing performance in tasks like semantic similarity and zero-shot classification~\cite{f170594b13efb4a93ef9819179fc929cac6809bf}.

Distinguishing between different types of uncertainty, such as epistemic (knowledge-based) and aleatoric (inherent randomness), has also been explored. Probing LLMs to identify when uncertainty arises from a lack of knowledge helps in assessing the reliability of their outputs~\cite{4862ea8e9bd568a7ea5d3347ebb6d2a8d7f80ecc}. Methods utilizing model explanations to measure confidence have been proposed; interpreting the consistency of generated explanations to quantify uncertainty~\cite{976f69e69a94535b88d6059550aa622ebb1c550f}. Efforts to improve the calibration of LLMs have revealed that while probabilities derived from model outputs are often miscalibrated, they can still predict correctness in tasks like multiple-choice question answering~\cite{fd18f1846c3e58dddd4054ad0cad90f140408ef1}. Comparing probabilistic confidence with verbalized confidence, studies have found that probabilistic measures are generally more accurate, though expressing internal confidence in natural language remains challenging for LLMs~\cite{bf184e23d67837e00bac4ec043c8afb691dc718c}. Techniques like reconfidencing LLMs to address grouping loss have been proposed to align confidence scores more closely with response accuracy~\cite{754f9c903754d909cb754364f4d6416ada0ab2b5}. Utilizing internal states from LLMs has shown promise in detecting hallucinations. By analyzing semantic information retained within models, methods like INSIDE leverage internal states for hallucination detection without relying solely on output tokens~\cite{f62acb5a743ea4d47a045460a9ee346c2cec5068}.

To evaluate uncertainty quantification methods comprehensively, datasets like MAQA have been introduced, focusing on data uncertainty arising from irreducible randomness~\cite{cf41f8d84004f34fd2cc7906ab6e99fff4e48f26}. Such resources enable assessment of UQ methods under realistic conditions. Advanced techniques have been developed to further refine uncertainty quantification in LLMs. For instance, conformal prediction methods have been applied to provide correctness coverage guarantees in open-ended natural language generation tasks, transforming heuristic uncertainty measures into rigorous prediction sets~\cite{bbc8eb04cbfa9f221dcd63d45ffd460b88a0ac01}. Supervised approaches leveraging labeled datasets have been proposed to estimate uncertainty in LLM responses, demonstrating benefits in enhancing uncertainty estimation and calibration across various tasks~\cite{0db2191b532a8ba6c4b89d016759fe6f7aa459b0}.

Information-theoretic metrics have been utilized to distinguish between epistemic and aleatoric uncertainties in LLM outputs, allowing for reliable detection of when the model's output is unreliable due to lack of knowledge~\cite{5322cd631e69ae484038b13ac320194afaccdc3b}. These approaches aid in understanding the limitations of LLMs and improving their trustworthiness in critical applications.

While significant progress has been made in uncertainty quantification for LLMs, there is a notable gap in research specifically addressing the use of entropy analysis for assessing GPT model uncertainty in mathematical OCR tasks. The recognition and extraction of mathematical expressions from images pose unique challenges due to the complexity of mathematical notation and the influence of image quality on recognition accuracy. Previous studies have not thoroughly explored how entropy measures can directly quantify uncertainty in GPT-based OCR systems handling mathematical content.

\textbf{Our Contribution.} In this paper, we aim to fill this gap by applying entropy analysis to assess the uncertainty of GPT models in the specific context of mathematical expression recognition from images. By calculating the conditional Shannon entropy of the output token sequences, we provide a quantitative measure of the model's uncertainty in generating LaTeX code corresponding to mathematical equations extracted from images at varying resolutions. This approach not only aligns with the theoretical foundations of information theory but also offers practical insights into optimizing OCR performance for complex mathematical content.

Our experimental results demonstrate a clear relationship between image resolution and GPT model uncertainty. Higher-resolution images result in lower entropy values, indicating reduced uncertainty and improved accuracy in the recognized LaTeX code. Conversely, lower-resolution images lead to increased entropy, reflecting higher uncertainty and a greater likelihood of recognition errors. These findings highlight the practical importance of considering image quality in GPT-based mathematical OCR applications and demonstrate how entropy analysis can effectively quantify model uncertainty in real-world tasks.

\textbf{Organization of the Paper.} The rest of the paper is organized as follows. In Section~\ref{InformationTheoryFundamentals}, we introduce fundamental concepts from information theory as they apply to OCR systems. We explore the mutual information between the input image and the recognized text \( I(X; Y) \) and discuss how recognition errors lead to information loss, as well as how effective OCR systems capture shared information despite noise. We model the OCR process using a noisy channel framework, analyzing how noise affects mutual information and the system's ability to accurately decode text.

In Section~\ref{ImpactofInputImageResolution}, we focus on the impact of input image resolution on the conditional entropy \( H(Y|X) \). Using the fact that the entropy of the output text \( H(Y) \) is (approximately) constant for a given text, we employ information-theoretic identities and inequalities to analyze how image resolution influences OCR uncertainties. We demonstrate that high-resolution images increase mutual information \( I(X; Y) \) and decrease conditional entropy \( H(Y|X) \), leading to improved recognition accuracy. Conversely, low-resolution images reduce mutual information and increase conditional entropy, resulting in higher uncertainty and error rates. We utilize the Data Processing Inequality and Fano's Inequality to support our analysis, providing a mathematical foundation for the observed effects of image resolution on OCR performance.

Section~\ref{GPTasEntropyModel} focuses on modeling GPT-based OCR systems using entropy measures. We describe how the GPT model processes the input, which includes both the image and a verbal prompt, to generate the output token sequence \( \boldsymbol{y} \). We calculate the conditional Shannon entropy \( H(\boldsymbol{y}|\boldsymbol{x}) \) of the output token sequence given the input, offering insights into the model's confidence and uncertainty during the text generation process. We introduce the concept of normalized entropy to facilitate comparisons across different sequences and resolutions. This section also discusses the practical computation of entropy using the GPT model's output probabilities and relates empirical findings to the theoretical concepts of conditional entropy \( H(Y|X) \) discussed earlier.

In Section~\ref{ExperimentalResults}, we present our experimental results obtained with the OpenAI GPT-4o model~\cite{openai_gpt_docs, openai_gpt4_report} for recognizing and extracting mathematical expressions from images at varying resolutions. We detail our experimental setup, including the selection of input data and configuration of the GPT-4o model with specific prompts and entropy calculations. We demonstrate a clear relationship between image resolution and the model's performance: as the image resolution decreases, the conditional entropy \( H(\boldsymbol{y}|\boldsymbol{x}) \) and the normalized conditional entropy increase, indicating higher uncertainty and a greater likelihood of recognition errors.

Section~\ref{LimitationFutureWork} discusses the limitations of our results and potential directions for future work. Finally, Section~\ref{Conclusion} provides a brief conclusion summarizing our findings and their implications.

\section{Information Theory Fundamentals of OCR}\label{InformationTheoryFundamentals}
\subsection{Entropy of Image and Recognized Text}

In OCR, the input image~\( X \) containing text is considered a source of information. The OCR system's output is the recognized textual information~\( Y \) extracted from the image. In real-world OCR systems, recognition errors (e.g., substituting a character due to noise) lead to some loss of information or misinterpretation, which reduces the accuracy of the recognition.

The entropy of the input image~\( H(X) \) is defined as:

\begin{equation}
H(X) = - \sum_{x \in \mathcal{X}} P(x) \log P(x),
\end{equation}
where~\( \mathcal{X} \) represents the set of possible pixel intensity values or features in the image, and \( P(x) \) is the probability distribution of these pixel intensities or features.

The entropy of the recognized text~\( H(Y) \) is defined as:

\begin{equation}
H(Y) = - \sum_{y \in \mathcal{Y}} P(y) \log P(y),
\end{equation}
where~\( \mathcal{Y} \) represents the set of possible recognized text outputs, and \( P(y) \) is the probability distribution over these outputs.

The joint entropy of the input image and the recognized text~\( H(X, Y) \) is defined as:

\begin{equation}
H(X, Y) = - \sum_{x \in \mathcal{X}} \sum_{y \in \mathcal{Y}} P(x, y) \log P(x, y),
\end{equation}
where~\( P(x, y) \) is the joint probability distribution of the input image features \( x \) and the recognized text outputs \( y \).

\subsection{Mutual Information Between Image and Recognized Text}
The mutual information~\( I(X; Y) \) between the input image~\( X \) and the recognized text~\( Y \) is a measure of the shared information between these two variables. It quantifies how much information about the text in the image is captured in the OCR output. Mutual information is calculated as:

\begin{equation}
I(X; Y) = H(X) + H(Y) - H(X, Y).
\end{equation}

High mutual information indicates that the OCR system has effectively captured the relevant information from the image, meaning that the image and text share a significant portion of information despite the potential presence of noise.

\subsection{Noise and Channel Model in OCR}\label{2.4.}

In OCR , input image can be thought about as being transmitted through a noisy channel (that is , variations in lighting , image resolution , or presence of artifacts ) to reach the OCR system , which decodes it into text. The entropy from noise~\( H(\text{Noise}) \) affects the system `s accuracy. In information theory terms : noise reduces mutual information~\( I(X; Y) \), as it actually creates ambiguity in the recognition process. From  channel model perspective, we have the following:
\begin{itemize}
    \item Noiseless channels would result in high mutual information, where~\( I(X; Y) \approx H(X) \), meaning all information in the image is retained in the recognized text.
    \item Noisy  channels have lower mutual information due to the increased uncertainty , reducing the effectiveness of the OCR.
\end{itemize}

The OCR system aims to maximize mutual information between an input and output by enhancing feature extraction ( e.g. , identifying edges of characters, reducing noise effects , and improving algorithms for  interpreting patterns as characters .

\subsection{Information Gain in OCR Processing}\label{2.5.}

When the OCR system successfully interprets the input image, the information gain in terms from mutual information \( I(X; Y) \) represents the effective amount of uncertainty reduction achieved by the recognition process. This gain corresponds to the OCR system ’s ability to distinguish characters from the background and noise , providing a measure of OCR accuracy and efficiency.

In summary,
\begin{itemize}
\item Mutual information \( I(X; Y) \) measures how much of text information is successfully extracted by OCR.
\item Noise affects the channel impacting the OCR system ’s ability to capture mutual information accurately.
\item A good OCR system maximizes mutual information , effectively reducing uncertainty in its output , thereby producing accurate text .
\end{itemize}
\section{Impact of Input Image Resolution on Conditional Entropy in OCR Systems}\label{ImpactofInputImageResolution}

In this section , we analyze how the resolution of input images affects the conditional entropy~\( H(Y|X) \) in Optical Character Recognition systems from the information - theoretic perspective , utilizing fact that the entropy of the output text~\( H(Y) \) is approximately constant for the given text. We employ fundamental identities and inequalities from information theory to  obtain the relationship between the image resolution and uncertainties involved in the OCR process.

\subsection{Entropy and Mutual Information Interpretation in OCR}\label{3.1.}
As we discussed in Section~\ref{InformationTheoryFundamentals}, in an OCR process , we model~\( X \) as the random variable representing the input image and~\( Y \) as the random variable representing the recognized text output . Entropy~\( H(Y) \) measures the average uncertainty or information content from the output text~\( Y \). In our context , since the text to be recognized is fixed ( i.e., we have a specific text in the image ) , the entropy~\( H(Y) \) is constant.

Conditional entropy~\( H(Y|X) \) quantifies a uncertainty remaining about \( Y \) after observing \( X \), representing the ambiguity in OCR system `s recognition about the text given the image. Mutual information~\( I(X; Y) \) quantifies amount about information that~\( X \) provides about \( Y \), representing the reduction in uncertainty of~\( Y \) due to the knowledge of~\( X \).

These quantities are related by the following identity:
\begin{equation}
I(X; Y) = H(Y) - H(Y|X)
\end{equation}

Given that~\( H(Y) \) is constant for the given text , any changes in \( H(Y|X) \) are directly due to changes in \( I(X; Y) \). Therefore , analyzing how the resolution affects~\( I(X; Y) \) allows us to understand its impact on~\( H(Y|X) \).

\subsection{Resolution Effects on Mutual Information and Conditional Entropy}\label{3.2.}

\subsubsection{High-Resolution Images}

High-resolution images capture detailed features about text characters enhancing OCR system `s ability to distinguish between similar characters. These increased details affect mutual  information~\( I(X; Y) \) and    conditional entropy~\( H(Y|X) \) as follows:
\begin{itemize}
    \item Mutual Information Increases: a detailed input image~\( X \) provides more specific information about the output text~\( Y \), thereby increasing~ \( I(X; Y) \):

  \begin{equation}
I_{\text{high-res}}(X; Y) > I_{\text{low-res}}(X; Y)
\end{equation}

\item Conditional Entropy Decreases: Since \( H(Y) \) is constant , an increase in \( I(X; Y) \) directly leads toward a decrease in \( H(Y|X) \):

\begin{equation}
H_{\text{high-res}}(Y|X) = H(Y) - I_{\text{high-res}}(X; Y) < H_{\text{low-res}}(Y|X)
\end{equation}

\end{itemize}

This implies that through high-resolution images the OCR system has less uncertainty in recognizing the text , leading to improved accuracy.

\subsubsection{Low-Resolution Images}

Low - resolution images provide fewer details, increasing ambiguity in character recognition. This affects~\( I(X; Y) \) and~\( H(Y|X) \) as follows:
\begin{itemize}
\item Mutual Information Decreases: the input image~\( X \) conveys less information about~\( Y \), reducing \( I(X; Y) \):

\begin{equation}
I_{\text{low-res}}(X; Y) < I_{\text{high-res}}(X; Y)
\end{equation}

\item  Conditional Entropy Increases: with~\( H(Y) \) being constant, a decrease in~\( I(X; Y) \) leads to an increase in~\( H(Y|X) \):

\begin{equation}
H_{\text{low-res}}(Y|X) = H(Y) - I_{\text{low-res}}(X; Y) > H_{\text{high-res}}(Y|X)
\end{equation}
\end{itemize}

Therefore, low-resolution images result in higher uncertainty by OCR system when recognizing the text , leading to a higher likelihood of errors .

\subsubsection{Data Processing Inequality}

Data Processing Inequality states that for any processing of  data, the following inequality holds:

\begin{equation}
I(X; Y) \geq I(f(X); Y),
\end{equation}
where~\( f(X) \) is an function about~\( X \). In the context of resolution reduction ; lowering the resolution can be viewed as applying a function~\( f \) that reduces the information content of~\( X \). Therefore, reducing resolution can not increase mutual information :

\begin{equation}
I_{\text{low-res}}(X; Y) = I(f(X); Y) \leq I(X; Y)
\end{equation}

This inequality  implies that lowering the resolution of the input image can only  reduce the OCR system `s knowledge about~\( Y \); thus, reducing~\( I(X; Y) \) and increasing~\( H(Y|X) \).

\subsubsection{Fano's Inequality}

Fano 's Inequality provides an upper bound on the probability of error~\( P_e \) in the channel:
\begin{equation}
H(Y|X) \geq H(P_e) + P_e \log(|\mathcal{Y}| - 1),
\end{equation}
where~\( H(P_e) = -P_e \log P_e - (1 - P_e) \log (1 - P_e) \) is the binary entropy function,
and~\( |\mathcal{Y}| \) is the size of the output alphabet (the number of possible characters). In the OCR context,~\( P_e \) is the error in character recognition. We observe that the lower~\( H(Y|X) \) implies the lower error rates~\( P_e \).

\subsection{Practical Implications for OCR Systems}\label{3.5.}

Optimizing the resolution of input images is crucial for minimizing~\( H(Y|X) \) and, therefore, reducing recognition errors :

- Higher Resolution Improves Accuracy: by increasing~\( I(X; Y) \) through higher resolution , we decrease~\( H(Y|X) \), thus, lowering the uncertainty in recognizing the text and reducing the probability of error as per Fano `s Inequality.

- Lower Resolution Degrades Performance : decreasing~\( I(X; Y) \) via lower resolution increases~\( H(Y|X) \), thus, raising the uncertainty and error rates.

\section{GPT as the Entropy Model}\label{GPTasEntropyModel}

In this section , we consider using a GPT model for task of recognizing and extracting mathematical expressions from a given image, and saving them in LaTeX format. In such a setting, the interaction with the GPT model is as  follows : the GPT model is provided with an input~\(\boldsymbol{x}\), which includes the image along with a verbal prompt, instructing the model to recognize math expressions in the image and output them in LaTeX format.

The model 's output is an text sequence~\(\boldsymbol{y} = (y_1, y_2, \ldots, y_n)\) composed from tokens. Let~\(\phi\) represent the GPT model , where~\(\phi(\boldsymbol{x}, y_{1:i-1}) = P_i(y_i \mid \boldsymbol{x}, y_1, y_2, \ldots, y_{i-1})\) models the probability distribution from the next token~\(y_i\) given the input \(\boldsymbol{x}\) and the preceding output tokens~\(y_1, y_2, \ldots, y_{i-1}\). The function~\(\phi(\boldsymbol{x}, \boldsymbol{y})\) outputs the sequence of next token probability distributions~\((P_1, P_2, \ldots, P_n)\) for the entire output sequence .

Thus, for a given input~\(\boldsymbol{x}\), the GPT model generates the output~\(\boldsymbol{y}\) along with auxiliary information consisting of the conditional probabilities~\(P_i(y_i \mid \boldsymbol{x}, y_{1:i-1})\) by each output token~\(y_i\).

\subsection{Shannon Entropy of the Output Token Sequence Given the Input}\label{4.1.}

The Shannon entropy measures the average amount of information or uncertainty associated with a random variable. In our context, for each position~$i$ in the output token sequence, the entropy~$H_i$ of the next-token distribution~$P_i$ is given by:
\begin{equation}
H_i = -\sum_{y \in \mathcal{V}} P_i(y \mid \boldsymbol{x}, y_{1:i-1}) \log P_i(y \mid \boldsymbol{x}, y_{1:i-1}),
\label{H_i}
\end{equation}
where~\(\mathcal{V}\) is the vocabulary of possible tokens.

The total conditional entropy~$H(\boldsymbol{y} | \boldsymbol{x})$ of the output token sequence~$\boldsymbol{y}$, given the input $\boldsymbol{x}$, is computed by summing the entropies at each position:
\begin{equation}
H(\boldsymbol{y} \mid \boldsymbol{x}) = \sum_{i=1}^n H_i = -\sum_{i=1}^n \sum_{y \in \mathcal{V}} P_i(y \mid \boldsymbol{x}, y_{1:i-1}) \log P_i(y \mid \boldsymbol{x}, y_{1:i-1}).
\end{equation}
This total conditional entropy represents the cumulative uncertainty of the GPT model in generating the output sequence~$\boldsymbol{y}$ given the input $\boldsymbol{x}$.

\subsection{Normalized Shannon Entropy of the Output Token Sequence}\label{4.2.}

To facilitate comparison across different sequences or models, we compute the normalized Shannon entropy at each position by dividing the entropy $H_i$ by the maximum possible entropy at that position. The maximum entropy occurs when the next-token distribution is uniform over the vocabulary~$\mathcal{V}$, in which case, we have~$H_i^{\text{max}} = \log |\mathcal{V}|$, where~$|\mathcal{V}|$ is the size of the vocabulary.

The normalized entropy~$h_i$ at position~$i$ is then defined as follows:
\begin{equation}
h_i = \frac{H_i}{H_i^{\text{max}}} = \frac{H_i}{\log |\mathcal{V}|}.
\end{equation}

Similarly, the normalized total entropy~$H(\boldsymbol{y} | \boldsymbol{x})$ of the output sequence~$\boldsymbol{y}$, given the input~$\boldsymbol{x}$, is defined as:
\begin{equation}
H(\boldsymbol{y} | \boldsymbol{x}) = \frac{H(\boldsymbol{y} | \boldsymbol{x})}{H^{\text{max}}(\boldsymbol{y})} = \frac{\sum_{i=1}^n H_i}{n \log |\mathcal{V}|} = \frac{1}{n} \sum_{i=1}^n h_i.
\end{equation}

The normalized entropy~$H(\boldsymbol{y} | \boldsymbol{x})$ ranges from~0 to~1, where~0 indicates no uncertainty (the model is certain about its next token at each position), and~1 indicates maximum uncertainty (the model's predictions are uniformly distributed over the vocabulary).

\subsection{Interpretation and Relevance}\label{4.3.}

Calculating the conditional Shannon entropy $H(\boldsymbol{y} | \boldsymbol{x})$ of the output token sequence provides insights into the GPT model's confidence and uncertainty during the generation process given the input $\boldsymbol{x}$. Lower entropy values suggest that the model is more confident about its predictions, which is desirable in tasks like mathematical expression recognition where precision is critical.

In the context of extracting mathematical expressions in LaTeX format, \emph{low entropy} indicates that the model confidently predicts the next token given the input, suggesting a high likelihood of correctly recognizing and formatting the mathematical expressions. \emph{High entropy} suggests uncertainty in the model's predictions given the input, which may lead to errors in the recognized expressions or incorrect LaTeX syntax.

By analyzing the entropy values, we assess the model's performance and identify positions in the output sequence, where the model may require additional support or where errors are more likely to occur.

\subsection{Practical Computation}\label{4.4.}
When the GPT model generates the output sequence~$\boldsymbol{y}$, it provides the conditional probabilities~$P_i(y_i \mid \boldsymbol{x}, y_{1:i-1})$ for each token $y_i$. We compute the entropy~$H_i$ at each position using these probabilities. If the full distribution~$P_i(y_i \mid \boldsymbol{x}, y_{1:i-1})$  is available (i.e. for each value of~$y \in \mathcal{V}$), we can compute~$H_i$ exactly as per equation~\eqref{H_i}. In other words, the exact calculation of the  entropy  requires access to the entire probability distribution over the vocabulary at each position.

In practice, however, the model only outputs the probabilities for a subset of tokens (e.g., the top-$k$ most probable tokens), so we approximate the entropy by considering only the available probabilities as follows:
\begin{equation}
H_i = - \sum_{y \in \tilde{\mathcal{V}}} P_i(y \mid \boldsymbol{x}, y_{1:i-1}) \log P_i(y \mid \boldsymbol{x}, y_{1:i-1}),
\end{equation}
where~$\tilde{\mathcal{V}}$ is a subset of the vocabulary of all possible tokens, such as, for example, the top-$k$ most probable tokens.

\subsection{Example Calculation}\label{4.5.}

Suppose the GPT model provides the following next-token probabilities (for simplicity, we consider a small subset of the vocabulary):

\begin{table}[h!]
\centering
\begin{tabular}{|c|c|}
\hline
Token~$y$ & $P_i(y \mid \boldsymbol{x}, y_{1:i-1})$ \\ \hline
`\textbackslash\{\}` & 0.60 \\ \hline
`\{` & 0.20 \\ \hline
`\}` & 0.10 \\ \hline
Other & Remaining probabilities summing to 0.10 \\ \hline
\end{tabular}
\end{table}
The entropy at position~\(i\) is calculated as follows:

\begin{equation}
\begin{aligned}
H_i &= - \left[ 0.60 \log 0.60 + 0.20 \log 0.20 + 0.10 \log 0.10 + 0.10 \log 0.10 \right] \\
&= - \left[ -0.442 + -0.322 + -0.230 + -0.230 \right] \\
&= 1.224 \text{ bits}
\end{aligned}
\end{equation}

Assuming a vocabulary size of \(|\mathcal{V}| = 10,000\), the maximum entropy at position \(i\) is:

\begin{equation}
H_i^{\text{max}} = \log 10,000 = 13.29 \text{ bits}
\end{equation}

The normalized entropy is:

\begin{equation}
h_i = \frac{H_i}{H_i^{\text{max}}} = \frac{1.224}{13.29} \approx 0.092
\end{equation}

A normalized entropy of approximately 0.092 indicates low uncertainty at this position given the input \(\boldsymbol{x}\).

\subsection{Relationship of~\( H(\boldsymbol{y} | \boldsymbol{x}) \) to~\( H(Y|X) \) }\label{4.8.}

The calculated conditional entropy~\( H(\boldsymbol{y} | \boldsymbol{x}) \) in this section provides a practical, empirical measure of the uncertainty the GPT model has in generating the output token sequence~\( \boldsymbol{y} \) given the input~\( \boldsymbol{x} \). This entropy relates directly to the theoretical concepts of conditional entropy~\( H(Y|X) \) discussed in Section~\ref{ImpactofInputImageResolution} as follows:
\begin{itemize}
\item Empirical Measurement of~\( H(Y|X) \): The entropy \( H(\boldsymbol{y} | \boldsymbol{x}) \) calculated here represents the sum of the uncertainties at each step in the generation of \( \boldsymbol{y} \) by the GPT model, given the input \( \boldsymbol{x} \). It serves as an empirical approximation of the conditional entropy \( H(Y|X) \) in the OCR context, where \( Y \) is the output text and \( X \) is the input image.

\item Connection to OCR Uncertainty: In Section~\ref{ImpactofInputImageResolution}, \( H(Y|X) \) quantifies the uncertainty in recognizing the text~\( Y \) given the image~\( X \). Similarly, \( H(\boldsymbol{y} | \boldsymbol{x}) \) reflects the GPT model`s uncertainty in generating each token~\( y_i \) given~\( \boldsymbol{x} \) (which includes the image and the prompt).

\item Impact of Image Resolution: The discussion in Section~\ref{ImpactofInputImageResolution} about how image resolution affects~\( H(Y|X) \) are mirrored in how the quality and clarity of the input image~\( \boldsymbol{x} \) influence~\( H(\boldsymbol{y} | \boldsymbol{x}) \). A higher-resolution image lead to lower entropy~\( H(\boldsymbol{y} | \boldsymbol{x}) \), indicating that the model has less uncertainty in generating the correct LaTeX code for the mathematical expressions.

\end{itemize}

\section{Experimental Results}\label{ExperimentalResults}

\subsection{Experimental Setup}

In this section, we present the results of our experiments using the OpenAI GPT-4o model~\cite{openai_gpt_docs, openai_gpt4_report} for recognizing and extracting mathematical expressions from images at different resolutions. The objective is to investigate how image resolution affects the model's performance, particularly in terms of entropy and uncertainty of the generated LaTeX code.

\subsubsection{Input Data}

We utilized the PDF document\cite{Zhou2021QuantumDC}. It contains complex mathematical expressions typical of research papers. The PDF page was converted to JPEG format at four different resolutions: 72, 96, 150, and 300 dots per inch (dpi).

\subsubsection{The GPT-4o Model Configuration}

For each of the four image resolutions, we interacted with the OpenAI GPT-4o model using specific prompts designed to instruct the model to recognize and convert mathematical equations into LaTeX format. The interaction was facilitated by the following Python code:

\begin{lstlisting}

system_prompt = """
You are an advanced Optical Character Recognition (OCR) system
designed specifically for mathematical content, integrated with
LaTeX conversion and verification capabilities.
Your primary tasks are:

1. **Accurately recognize mathematical equations** from provided
 images of research paper pages. This includes handling complex
  mathematical symbols, operators, fractions, integrals, summations,
  subscripts, superscripts, and other mathematical notations.

2. **Convert the recognized equations** into syntactically correct
 and semantically accurate LaTeX code. "Semantically accurate" means
  that the LaTeX code must exactly match the original equation
   in terms of symbols, structure, and formatting
    as it is visually represented.

3. **Provide the recognized equations
 in the following LaTeX output format**:
```
\begin{equation}
[LaTeX code of the equation] \tag{equation number if present}
\end{equation}
```
- Ensure that each equation is formatted properly
 and includes the correct LaTeX syntax.
- The "equation number" should be included only
 if it appears in the image.

4. **In your response**, provide only the LaTeX code in
 the specified format without any explanations or comments.
"""

user_prompt = """
Process the attached image of a research paper page and extract
 all mathematical equations present.
Provide only the LaTeX code in the following output format.
"""

messages = [
    {"role": "system", "content": system_prompt},
    {
        "role": "user",
        "content": user_prompt,
        "attachments": [
            {
                "type": "image",
                "url": "https://myserver/2106.13823v3-P3-300.jpg"
            }
        ]
    }
]

completion = client.chat.completions.create(
    model="gpt-4o",
    messages=messages,
    seed=12345,
    temperature=0,
    logprobs=True,
)

\end{lstlisting}

For technical reasons, the verbal instructions to the model are provided via two prompts: \texttt{system\_prompt} and \texttt{user\_prompt}. \texttt{system\_prompt} defines the model's role and tasks, while \texttt{user\_prompt} provides specific instructions for processing the image.

\subsubsection{Entropy Calculation}
When calling the GPT model,  we enabled the `logprobs` option to obtain the log probabilities of each token generated by the model. This allowed us to compute the (conditional) entropy of the output token sequence, serving as a measure of the model's uncertainty at each step.

For each output token \( y_i \), we calculated the negative log probability:

\begin{equation}
H_i = -\log_2 P_i(y_i \mid \boldsymbol{x}, y_{1:i-1})
\end{equation}

The total conditional entropy \( H(\boldsymbol{y} \mid \boldsymbol{x}) \) is the sum of the individual entropies:

\begin{equation}
H(\boldsymbol{y} \mid \boldsymbol{x}) = \sum_{i=1}^n H_i
\end{equation}

We also computed the normalized entropy by dividing the total entropy by the maximum possible entropy for the sequence, which is \( n \log_2 |\mathcal{V}| \), where \( n \) is the number of tokens and \( |\mathcal{V}| \) is the vocabulary size.


\subsubsection{Experiment 1: 300 dpi Resolution}

Original document: [arXiv:2106.13823] Page-3

Image file: 2106.13823v3-P3-300.jpg

Image resolution: 300 dpi

Number of output tokens: 453

Total Entropy: 7.77 bits

Normalized Entropy: 0.0171 bits

\parskip=2pt
\noindent  \underline{\textbf{Recognized Math Equations:}}

\begin{equation}
Q(N, \rho, \epsilon) = \sum_{x \in \epsilon\text{-strong-typical}} |i_1\rangle \langle i_1| \otimes |i_2\rangle \langle i_2| \otimes \cdots \otimes |i_N\rangle \langle i_N|. \tag{11}
\end{equation}

\begin{equation}
\text{tr} \left( Q(N, \rho, \epsilon) \rho^{\otimes N} \right) \geq 1 - \delta. \tag{12}
\end{equation}

\begin{equation}
\text{tr} \left( Q(N, \rho, \epsilon) \rho^{\otimes N} \right) = \sum_{x \in \epsilon\text{-strong-typical}} P(i_1) P(i_2) \cdots P(i_N) = \sum_{x \in \epsilon\text{-strong-typical}} P(x). \tag{13}
\end{equation}

\begin{equation}
\langle l \rangle = \sum_{i=1}^{D} p_i l_i = \sum_{i=1}^{D} p_i \log \frac{1}{q_i} = H(p, q), \tag{16}
\end{equation}

\begin{equation}
H(p, q) \leq \langle l \rangle < H(p, q) + 1. \tag{17}
\end{equation}

\begin{equation}
\sigma = \sum_{i=1}^{D} q_i |a_i\rangle \langle a_i|, \quad \rho = \sum_{i=1}^{D} p_i |b_i\rangle \langle b_i|, \quad \{q_i\} \neq \{p_i\}, \quad \{|a_i\rangle\} \neq \{|b_i\rangle\}. \tag{15}
\end{equation}


\subsubsection{Experiment 2: 150 dpi Resolution}

Original document: [arXiv:2106.13823] Page-3

Image file: 2106.13823v3-P3-150.jpg

Image resolution: 150 dpi

Number of output tokens: 434

Total Entropy: 12.28 bits

Normalized Entropy: 0.028 bits;

\parskip=2pt
\noindent  \underline{\textbf{Recognized Math Equations:}}

\begin{equation}
Q(N, \rho, \epsilon) = \sum_{|i\rangle \in \epsilon\text{-strong-typical}} |i_1\rangle \langle i_1| \otimes |i_2\rangle \langle i_2| \otimes \cdots \otimes |i_N\rangle \langle i_N|. \tag{11}
\end{equation}

\begin{equation}
\text{tr} \left( Q(N, \rho, \epsilon) \rho^{\otimes N} \right) \geq 1 - \delta. \tag{12}
\end{equation}

\begin{equation}
\text{tr} \left( Q(N, \rho, \epsilon) \rho^{\otimes N} \right) = \sum_{x \in \epsilon\text{-strong-typical}} P(i_1) P(i_2) \cdots P(i_N) = \sum_{x \in \epsilon\text{-typical}} P(x). \tag{13}
\end{equation}

\begin{equation}
H(p, q) = \sum_{i=1}^{D} p_i \log \frac{1}{q_i} = H(q, p), \tag{16}
\end{equation}

\begin{equation}
H(p, q) \leq \langle l \rangle < H(p, q) + 1. \tag{17}
\end{equation}

\begin{equation}
\sigma = \sum_{i=1}^{D} q_i |a_i\rangle \langle a_i|, \quad \rho = \sum_{i=1}^{D} p_i |b_i\rangle \langle b_i|, \quad \{a_i\} \neq \{p_i\}, \quad \{a_i\} \neq \{b_i\}. \tag{15}
\end{equation}


\subsubsection{Experiment 3: 96 dpi Resolution}

Original document: [arXiv:2106.13823] Page-3

Image file: 2106.13823v3-P3-96.jpg

Image resolution: 96 dpi

Number of output tokens: 476

Total Entropy: 18.27 bits

Normalized Entropy: 0.038 bits

\parskip=2pt
\noindent  \underline{\textbf{Recognized Math Equations:}}

\begin{equation}
Q(N, \rho, \epsilon) = \sum_{|i\rangle \, \epsilon\text{-strongly typical}} |i\rangle \langle i| \otimes |i_2\rangle \langle i_2| \otimes |i_N\rangle \langle i_N|      \tag{11}
\end{equation}

\begin{equation}
\text{tr} \left( Q(N, \rho, \epsilon) \rho^{\otimes N} \right) \geq 1 - \delta  \tag{12}
\end{equation}

\begin{equation}
\text{tr} \left( Q(N, \rho, \epsilon) \rho^{\otimes N} \right) = \sum_{x \, \epsilon\text{-strongly typical}} P(i_1) P(i_2) \cdots P(i_N) = \sum_{x \, \epsilon\text{-strongly typical}} P(x)   \tag{13}
\end{equation}

\begin{equation}
(1 - \delta)^2 2^{N(S(\rho) - \epsilon)} \leq T(N, \rho, \epsilon) \leq 2^{N(S(\rho) + \epsilon)}       \tag{14}
\end{equation}

\begin{equation}
\sigma = \sum_{i=1}^{D} q_i |a_i\rangle \langle a_i|, \quad \rho = \sum_{i=1}^{D} p_i |b_i\rangle \langle b_i|, \quad \{q_i\} \neq \{p_i\}, \quad \{|a_i\rangle\} \neq \{|b_i\rangle\}  \tag{15}
\end{equation}

\begin{equation}
l(x) = \left\lceil \log \frac{1}{q_x} \right\rceil = H(p, q)    \tag{16}
\end{equation}

\begin{equation}
H(p, q) \leq l(x) < H(p, q) + 1         \tag{17}
\end{equation}


\subsubsection{Experiment 4: 72 dpi Resolution}

Original document: [arXiv:2106.13823] Page-3

Image file:  2106.13823v3-P3-72.jpg

Image resolution: 72 dpi

Number of output tokens: 473

Total Entropy: 38.91 bits

Normalized Entropy: 0.082 bits

\parskip=2pt
\noindent  \underline{\textbf{Recognized Math Equations:}}

\begin{equation}
Q(\mathcal{N}, \rho, \epsilon) = \sum_{k: \langle k | \rho | k \rangle \geq \epsilon} |k\rangle \langle k| \otimes |k\rangle \langle k| \otimes |k\rangle \langle k|    \tag{11}
\end{equation}

\begin{equation}
\operatorname{tr} (Q(\mathcal{N}, \rho, \epsilon) \rho^{\otimes n}) \geq 1 - \delta     \tag{12}
\end{equation}

\begin{equation}
\operatorname{tr} (Q(\mathcal{N}, \rho, \epsilon) \rho^{\otimes n}) = \sum_{x \in \epsilon\text{-strong-typical}} P(x) \cdot P(x) \cdot \ldots \cdot P(x) = \sum_{x \in \epsilon\text{-strong-typical}} P(x)    \tag{13}
\end{equation}

\begin{equation}
(1 - \delta) 2^{n(S(\rho) - \epsilon)} \leq P(\mathcal{N}, \rho, \epsilon) \leq 2^{n(S(\rho) + \epsilon)}       \tag{14}
\end{equation}

\begin{equation}
\langle \ell \rangle = \sum_{x} P(x) \cdot \left\lceil \log \frac{1}{P(x)} \right\rceil = H(P, q)       \tag{16}
\end{equation}

\begin{equation}
H(q) \leq \langle \ell \rangle \leq H(q) + 1.   \tag{17}
\end{equation}

\begin{equation}
\sigma = \sum_{i} \lambda_{i} |\phi_{i}\rangle \langle \phi_{i}| = \sum_{i} p_{i} |\psi_{i}\rangle \langle \psi_{i}|    \tag{18}
\end{equation}

\subsection{Analysis and Discussion}\label{5.3.}

\begin{table}[h!]
\centering
\begin{tabular}{|p{2cm}|p{2cm}|p{2cm}|p{2cm}|p{2.5cm}|}

\hline
\textbf{Image Resolution (dpi)} & \textbf{Number of Output Tokens} & \textbf{Total Entropy (bits)} & \textbf{Normalized Entropy (bits)} &  \textbf{Recognition Accuracy} \\ \hline
300 & 453 & 7.77  & 0.0171 & Perfect \\ \hline
150 & 434 & 12.28 & 0.028  & Near perfect \\ \hline
96  & 476 & 18.27 & 0.038  & With some errors \\ \hline
72  & 473 & 38.91 & 0.082  & With many errors \\ \hline
\end{tabular}
\caption{Comparison of entropy values at different image resolutions.}
\label{tab:entropy_resolution}
\end{table}

The experimental  results demonstrate how image resolution affects the GPT model's performance in recognizing and converting mathematical expressions into LaTeX format.

- Entropy Trends: As the image resolution decreases, the total conditional entropy \( H(\boldsymbol{y} | \boldsymbol{x}) \) increases.  This indicates that the model's uncertainty in generating the output tokens grows when processing lower-resolution images.

- Normalized Entropy: The normalized entropy per token also increases with decreasing resolution, from 0.0171 bits per token at 300 dpi to 0.0823 bits per token at 72 dpi. This suggests that each token generated from lower-resolution images carries more uncertainty.

- Output Tokens: The number of output tokens remains relatively stable across different resolutions, with slight variations.

- Impact on Accuracy: Higher entropy values are associated with an increased likelihood of errors in the recognized mathematical expressions. Lower-resolution images provide less detailed information, leading to higher uncertainty in the model's predictions and recognition errors, in perfect agreement with the information-theoretic results discussed in  previous sections.

\section{Limitations and Future Work}\label{LimitationFutureWork}

- Entropy Calculation: The entropy calculations are based on the log probabilities provided by the model. If the model does not output the full probability distribution over the vocabulary, the entropy estimates may be approximations.

- Model Limitations: The OpenAi GPT-4o model's performance may be influenced by factors beyond image resolution, such as its training data and inherent capabilities in handling mathematical notation.

- Further Experiments: Future work could involve testing with a broader range of image qualities, different OCR and GPT models, and incorporating noise or distortions to simulate real-world scenarios.

- Error Analysis: A detailed analysis of the errors in the generated LaTeX code could provide insights into specific challenges faced by the model at different resolutions.

\section{Conclusion}\label{Conclusion}

In this paper, we investigated the uncertainty of GPT models in extracting mathematical equations from images of varying resolutions and converting them into LaTeX code. By employing concepts of entropy and mutual information, we examined the recognition process and assessed the model's uncertainty in this specialized OCR task. Our study aimed to fill a notable gap in the existing research by specifically addressing the use of entropy analysis for assessing GPT model uncertainty in mathematical OCR tasks.

Our experimental results demonstrate a clear relationship between input image resolution and the GPT model's uncertainty. Higher-resolution images lead to lower entropy values, indicating reduced uncertainty and improved accuracy in the recognized LaTeX code. Conversely, lower-resolution images result in increased entropy, reflecting higher uncertainty and a higher likelihood of recognition errors. These findings empirically validate the theoretical framework established in earlier sections, highlighting the connection between the  model's accuracy and the entropy of the output token sequence.

The  contributions of this work lie in connecting information-theoretic concepts, such as conditional entropy and mutual information, to the practical performance of GPT-based OCR systems. By calculating the conditional Shannon entropy of the output token sequences, we provided a quantitative measure of the model's uncertainty. This approach not only aligns with the theoretical foundations of information theory but also extends its application to modern LLMs in real-world tasks.

In conclusion, our work highlights the critical role of entropy analysis in quantifying GPT model uncertainty in mathematical OCR tasks. We believe that this study opens avenues for further research, encouraging the exploration of entropy-based methods in assessing and improving the performance of LLMs across various complex tasks.

\section*{Declarations}

\subsection*{Conflict of Interest}
The authors have no financial or non-financial interests to disclose. The authors have no Conflict of Interest to declare for the content of this article.

\subsection*{Data Availability}
The data used in this study consist of images derived from a publicly available PDF document on arXiv (\url{https://arxiv.org/abs/2106.13823}). The PDF page used in the experiments was converted to JPEG format at four different resolutions (72, 96, 150, and 300 dpi). These images are not included in the GitHub repository but can be recreated by accessing the publicly available paper and following the instructions provided in the repository.

\subsection*{Code Availability}
The Python code used in this study is openly available on GitHub at \url{https://github.com/Alexei-WLU/gpt-ocr-entropy-analysis}. The repository includes all scripts necessary to reproduce the experimental results presented in this paper. Note that the images derived from the arXiv PDF are not included in the repository. Instructions for obtaining the PDF and generating the images are provided in the repository's README file.


\bibliography{OCR_entropy.bib}

\end{document}